\documentclass[11 pt]{article}
\usepackage[dvips]{graphicx}

\input{tcilatex}
\begin{document}

\title{The wetting problem of fluids on solid surfaces. Part 1: the dynamics of
contact lines}
\author{Henri GOUIN\thanks{
E-mail:henri.gouin@univ.u-3mrs.fr, telephone: +33 491288407, fax:
+33 491288776}}
\date{Laboratoire de Mod{\'e}lisation en M{\'e}canique et
Thermodynamique, EA2596, Universit{\'e} d'Aix-Marseille, 13397
Marseille Cedex 20, France} \maketitle \centerline {{\it
Communicated by\ } Kolumban Hutter, {\it Darmstadt}}

\begin{abstract}
The understanding of the spreading of liquids on solid surfaces is
an important challenge for contemporary physics. Today, the motion
of the contact line formed at the intersection of two immiscible
fluids and a solid is still subject to dispute.\newline In this
paper, a new picture of the dynamics of wetting is offered through
an example of non-Newtonian slow liquid movements. The kinematics
of liquids at the contact line and equations of motion are
revisited. Adherence conditions are required except at the contact
line. Consequently, for each fluid, the velocity field is
multivalued at the contact line and generates an equivalent
concept of line friction but stresses and viscous dissipation
remain bounded. A Young-Dupr\'e equation for the apparent dynamic
contact angle between the interface and solid surface depending on
the movements of the fluid near the contact line is proposed.
\end{abstract}

\bigskip

\noindent \textit{Key words}: contact angle, contact line, dynamic
Young-Dupr\'{e} equation, wetting
\section{ Introduction}

The spreading of fluids on solid surfaces constitutes a
significant field of research into the processes met in nature,
biology and modern industry. Interfacial phenomena relating to
gas-liquid-solid systems take into account contact angles and
contact lines which are formed at the intersection of two
immiscible fluids and a solid. The interaction between the three
materials in the immediate vicinity of the contact line has great
effects on the statics and dynamics of flows (Dussan,
\cite{dussan2}). Many observations associated with the motion of
two fluids in contact with a solid wall were performed  (Bataille,
\cite{bat}; Dussan and Davis, \cite{dussan1}; Dussan, Ram\'e and
Garoff, \cite{dussan3}; Pomeau, \cite{pomeau}). According to the
advance or the recede of a fluid on a wall, we observe the
existence of an apparent dynamic contact angle when a contact line
is moving. This angle, named after Young, depends on the celerity
of the contact line, and the motion in the vicinity of the contact
line does not seem to be influenced by the behaviour of the total
flow (Bazhelakov and Chesters, \cite{baz}; Blake, Bracke and
Shikhmurzaev, \cite{blake2}). It is noteworthy that since Young's
article on capillarity, \cite{Young}, the understanding of these
phenomena has remained incomplete. For example, it is well known
that for Newtonian fluids the total dissipation and the interface
curvature at the contact line are infinite  (Huh and Scriven,
\cite{huh}; Dussan and Davis, \cite{dussan1}; Pukhnachev and
Solonnikov, \cite{puk}). In fact, fundamental questions remain
unanswered. Among these are the following:
\newline
What is the kinematics of the contact line? Can the fluid velocity
fields be multivalued on this line? What is the work of the
dissipative forces in its vicinity? Is there slip of the contact
line on the solid wall? What is the connection between apparent
and intrinsic contact angles? \newline There are various ways to
overcome these difficulties: to consider the slip length on the
solid wall (Hocking, \cite{hocking}; Shikhmurzaev, \cite{shik}),
to consider one phase as a perfect fluid, the possibility of a
thin film as a precursor film on a wall (de Gennes,
\cite{gennes}), the assumption of dynamic surface tension
different from the static counterpart (Shikhmurzaev, \cite{shik}),
the use of non-linear capillary theories such as Cahn and
Hilliard's theory of capillarity (Seppecher, \cite{seppecher}), or
the direct computation of flows by means of molecular models
(Koplick, Banavar and Willemsen, \cite{kop}). All these attempts
are not able to produce a complete satisfactory answer to the
previous questions.

\noindent It was noticed by using molecular methods that large
amplitude shearing rates reveal a tendency to reorganize the
liquid, to facilitate the flow and to reduce the viscosity. This
suggests that in reality there may be rheological anomalies around
the contact line (Heyes {\it et al},  \cite{heyes}; Holian and
Evans, \cite{Evans}; Ryckaert {\it et al}, \cite{ryckaert}).

\noindent For condensed matter and far from critical conditions,
interfaces which are transition layers of the size of a few
Angstr\"oms between fluids or between a fluid and a solid can be
modelled by surfaces endowed with a capillary energy (Rowlinson
and Widom, \cite{row}). Solid walls are rough on a molecular or
even microscopic scale. Moreover, the chemical inhomogeneity due
to the nature of the solid or the presence of surfactants changes
the surface tension in a drastic way. Nevertheless, roughness for
example is taken into account by corrections of the measurement on
a mean geometric surface (Wenzel equation in Cox, \cite{cox} or
Wolansky and Marmur, \cite{wol}).

\medskip

\noindent The motion of liquids in contact with a solid wall will
be considered in this paper within the framework of continuum
mechanics. Knowledge of the equations and boundary conditions
which govern the movements of liquids in contact with solid walls
and  control of the contact line motion are the aim of our
study.\newline We propose a model of the dynamics of wetting for
slow movements. To prove its accuracy, we are only considering
\textit{partial wetting} when the balance contact angle is
theoretically defined without ambiguity  (de Gennes,
Brochard-Wyart and Qu\'er\'e, \cite{brochard}). The liquids are
non-Newtonian; so the viscous stress tensor deviates from the
Navier-Stokes model for large values of the strain rate tensor.
For two-dimensional flows, and in the lubrication approximation,
the streamlines have an analytic representation and it is possible
to obtain the flows near the contact line. Equations of motion,
boundary conditions and some consequences on the contact angle
behaviour are deduced.

\bigskip The notation is that of ordinary Cartesian tensor
analysis (Serrin, \cite{serrin}). In a fixed coordinate system,
the components of a vector (covector) \ $\mathbf{a}$ are denoted
by $a^i $, ($a_i $), where $i = 1, 2, 3 $. In order to describe
the fluid motion analytically, we refer to the coordinates
$\mathbf{x} \equiv\ (x^1,x^2,x^3)$ as the particle's position
(Eulerian variables). The corresponding reference position is
denoted by $\mathbf{X} \equiv\ (X^1,X^2,X^3)$ (Lagrangian
variables). The motion of a fluid is classically represented by
the transformation $\displaystyle \mathbf{x} = {\mbox{{\boldmath
$\varphi$}}} (t,\mathbf{X}) \ \mathrm{or} \ x^i =
\varphi^i(t,\mathbf{X}) $ . It is assumed that ${\mbox{{\boldmath
$\varphi$}}} $ possesses an inverse $\mathbf{X} = \Phi(t,
\mathbf{x}) $ and continuous derivatives up to the second order
except at certain surfaces and curves. The vector $\mathbf{V}$
denotes the fluid velocity.
 The whole domain occupied by the fluid in Lagrangian variables is
 $D_0$ and its boundary  is the surface $\Sigma_0$. In Eulerian
variables, the fluid occupies the volume $D_t$ with boundary
$\Sigma_t$ corresponding to the fixed regions $D_0,\Sigma_0$ in
the reference configuration. A moving curve $\Gamma_t$ on
$\Sigma_t$ in the present configuration  corresponds to the moving
curve $\Gamma_{0t}$ on $\Sigma_0$ in the reference configuration.
The domains $D_0,D_t,\Sigma_0,\Sigma_t,\Gamma_{0t}, \Gamma_t$ must
obviously be oriented differentiable manifolds.

\noindent  To each point of $\Sigma_t$ a  unit normal vector
$\mathbf{n} $ ($n^i$), external to $D_t$, and a mean radius of
curvature, $R_m$ can be assigned. Furthermore $
\mathbf{Id} $ is the identity tensor with components $\delta_i^j$. Then $%
\mathbf{Id} - \mathbf{n \otimes n}$ (components $\delta_i^j - n^j
n_i$ ) is the projection operator onto the tangent plane of the
surface $\Sigma_t$;  let $\mathbf{t}$ denote the unit tangent
vector of $\Gamma_t$ oriented; $\mathbf{n^{\prime}} = \mathbf{n}
\times \mathbf{t}$ is the binormal vector to $\Gamma_t$ with
respect to $\Sigma_t$; it is a vector lying in the surface
$\Sigma_t$.

\section{General kinematics of a liquid at a contact line}

Following Dussan  and Davis' experiments for contact line
movements,  \cite{dussan1}, the usual stick-adhesive point of view
of fluid adherence at a solid wall is disqualified in continuum
mechanics. A liquid which does not slip on a solid surface does
not preclude the possibility that at some instant a liquid
material point may leave the surface. The no-slip condition is
expressed as follows:

\centerline {\it The velocity of the liquid must equal the solid
velocity at the surface.}
\begin{figure}[h]
\begin{center}
\includegraphics[width=10cm]{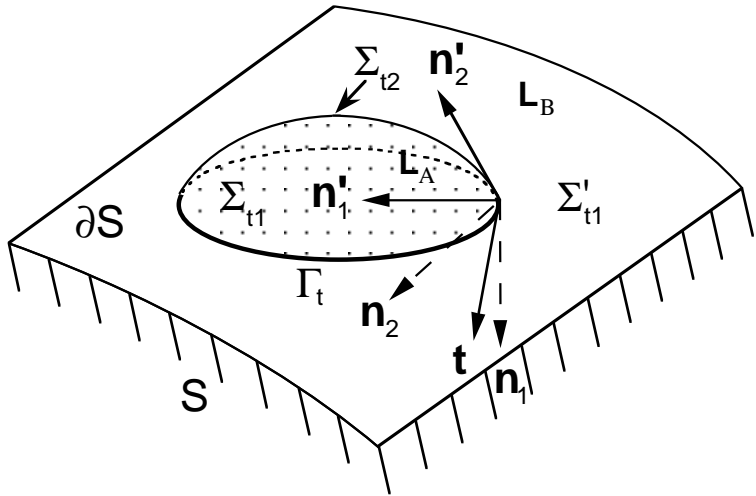}
\end{center}
\caption{A liquid $L_A$ (in drop form) lies on a solid surface
$\partial S$. The liquid $L_A$ is bordered with a fluid $L_B$ and
a solid $S$; $\Sigma_{t1}$ is the boundary between liquid $L_A$
and solid $S$; $ \Sigma^{\prime}_{t1}$ is the boundary between
fluid $L_B$ and solid $S$, and consequently, $ \Sigma_{t1} \cup
\Sigma^{\prime}_{t1} =
\partial S$;  $\Sigma_{t2}$ is the interface
between liquid $L_A$ and fluid $L_B$; $\mathbf{n}_1$ and
$\mathbf{n}_2$ are the unit normal vectors to $\Sigma_{t1}$ and
$\Sigma_{t2}$, exterior to the domain of liquid $L_A$ and the
domain of fluid $L_B$, respectively; the edge $\Gamma_t$ (or
contact line) is common to $\Sigma_{t1}$ and $\Sigma_{t2}$ and
$\mathbf{t}$ is the unit tangent vector to $\Gamma_t$ relative to
$\mathbf{n}_1$;
 $\mathbf{n^{\prime}}_1$ and  $\mathbf{n^{\prime}}_2$ are the
binormals to $\Gamma_t$ relative to $\Sigma_{t1}$ and
$\Sigma_{t2}$, respectively. }
\label{fig1}
\end{figure}

\noindent Let a liquid $L_A$ be in contact with $(i)$ a solid body
$S$ on an imprint $\Sigma_{t1}$ of the boundary $\partial S$ of
$S$ and $(ii)$ an incompressible fluid $L_B$ along an interface
$\Sigma_{t2}$ (fig. \ref {fig1}). Let, moreover, the mobile
surfaces $\Sigma_{t\alpha} \, (\alpha = 1, 2)$ be described by the
Cartesian equations $f_\alpha(t,\mathbf{x}) = 0 \, $ $(\alpha = 1,
2)$ and let the equations of the common curve $ \Gamma_t =
\Sigma_{t1} \cap \Sigma_{t2} $ be given by
\[
f_1(t,\mathbf{x}) = f_2(t,\mathbf{x}) = 0.
\]
\noindent For a geometric point M of $\Sigma_{t\alpha}$ with
velocity $ \mathbf{W}_\alpha$ we obtain the kinematic relation
\[
{\frac{\partial f_\alpha }{\partial x^i}} W_\alpha^i +
{\frac{\partial f_\alpha }{\partial t}} = 0,
\]
in which the usual convention that over a doubly repeated index
summation from $1$ to $3$ is understood. With the notations of
fig. \ref{fig1},  if we observe that $\displaystyle n_{\alpha i} =
\lambda_\alpha (\partial f_\alpha /\partial x^i)$ where
$\lambda_\alpha$ is a suitable scalar, the celerity of the surface
$\Sigma _{t\alpha}$ has the value $\displaystyle c_\alpha = {\
n_{\alpha i}} W_\alpha^i = - \lambda_\alpha (\partial f_\alpha
/\partial t) $. This celerity depends only on the coordinates $(t,
\mathbf{x})$ of M.

\noindent Let the velocity of a point of $\Gamma_t$ be denoted by
$\mathbf{W}$. We shall denote the unit tangent vector to
$\Gamma_t$ relative to $\mathbf{n}_1$   simply by $\mathbf{t}$ and
consequently, $\mathbf{n^{\prime}_1} = {\mathbf{n}}_1 \times
\mathbf{t}$, $\mathbf{n^{\prime}_2} = {\mathbf{n}}_2 \times
\mathbf{t}$ (see fig. 1). The velocity of the common line is then
expressible as
\[
\mathbf{u} = (\mathbf{Id} -\mathbf{t}\otimes\mathbf{t})
\mathbf{W}.
\]
It is orthogonal to $\Gamma_t$ and its expression depends only on
the coordinates $(t,\mathbf{x})$ of the point on $\Gamma_t$ but it
is not necessarily tangential to $\partial S$. Then, $\mathbf{u} =
\beta_1 \mathbf{n}^{\prime}_1 + \beta_2 \mathbf{n}^{\prime}_2$,
where $\beta_1$ and $\beta_2$ are two scalars. Thus, along
$\Gamma_t$, $n_{\alpha i} W^i = c_\alpha $ and $c_1 =
 n_{1 i} W^i =  n_{1 i} u^i = \beta_2 (\mathbf{n}_1,\mathbf{n}_2,
 \mathbf{t})$, $c_2 =
 n_{2 i} W^i =  n_{2 i} u^i = \beta_1 (\mathbf{n}_2,\mathbf{n}_1,
 \mathbf{t})$. Consequently,
\[
\mathbf{u} = {\frac{ c_2 \,\mathbf{n}^{\prime}_1 - c_1 \,\mathbf{n}%
^{\prime}_2  }{{(\mathbf{n}_1, \mathbf{t}, \mathbf{n}_2)}}},
\]
in which $(\mathbf{a},\mathbf{b}, \mathbf{c})$ is the triple
product of the three vectors $\mathbf{a},\mathbf{b}, \mathbf{c}$.
Due to the definitions of $\mathbf{n}_1$ and $\mathbf{n}_2$, we
remark that  $(\mathbf{n}_1, \mathbf{t}, \mathbf{n}_2) > 0$.

\noindent The kinematics of fluids in the vicinity of the contact
line will be axiomatized as follows:

\noindent $\Sigma_{t1}$ is a part of the surface of the solid $S$.
Liquid $ L_A $ adheres to $\partial S$ in the sense of the no-slip
condition previously proposed. $\Sigma_{t2}$ is a material surface
of liquid $ L_A $. \medskip

\noindent At the contact line, the velocity $\mathbf{u}$ may have
any direction between $\mathbf{n}_1$ and $\mathbf{n}_2$, depending
upon, how the contact line is approached within $L_A$. On the
solid surface  $\Sigma_{t1}$ of $S$ it is, however, tangential to
$\Sigma_{t1}$, so that
\[
\mathbf{u} = {\frac{c_2 \,\mathbf{n}^{\prime}_1 }{{(\mathbf{n}_1,
\mathbf{t}, \mathbf{n}_2)}}} \equiv {\frac{(n_{2 i} V_2^i )
\,\mathbf{n}^{\prime}_1 }{{ \ (\mathbf{n}_1, \mathbf{t},
\mathbf{n}_2)}}} \equiv - u \, \mathbf{n}^{\prime}_1
\]
where $u$ denotes the value of the contact line celerity in the
direction liquid $L_A$ to fluid $L_B$  and $\mathbf{V}_2$ is the
common velocity of the fluids on $ \Sigma _{t2}$. The contact line
$\Gamma _t$ is \textit{not a material line} of $ L_A $; its
velocity is different from the velocities of the liquid on $ S $
and on $\Sigma _{t2}$.

\begin{figure}[h]
\begin{center}
\includegraphics[width=10cm]{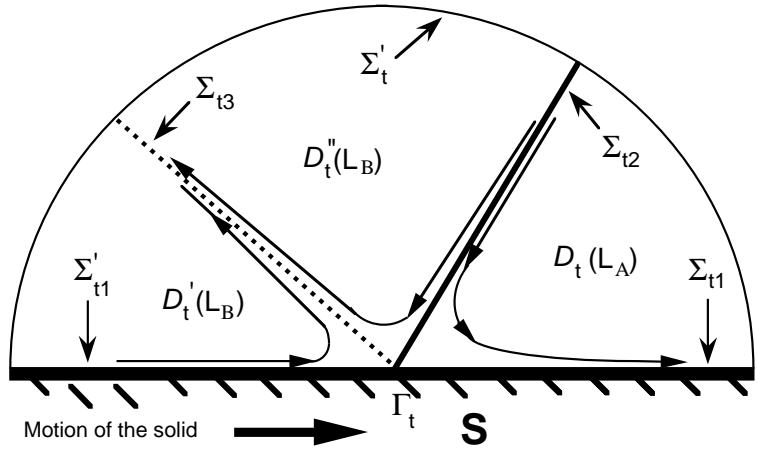}
\end{center}
\caption{Typical two-dimensional motion of fluids in contact on a
solid surface with a stationary contact line. The wedges formed by
$ \Sigma _{t1}$, $ \Sigma _{t2}$ and $ \Sigma _{t2}$,
$\Sigma^{\prime}_{t1}$ bound the fluids $L_A$ and $L_B$. The
auxiliary surface $ \Sigma _{t3}$ separates $L_B$ into two
domains. A control surface $ \Sigma^{\prime}_t$ together with $
\Sigma _{t1}$ and $\Sigma^{\prime}_{t1}$  constitute the boundary
of a compact domain $D_t$ of the two fluids. For explanations, see
main text.} \label{fig2}
\end{figure}

\noindent The motion of the particles of $ L_A$ on $\partial S$ and $%
\Sigma_{t2}$ is comparable with that of an adhesive tape stuck on
a wall, the other edge of the adhesive tape being mobile (fig.
\ref{fig2}): for $u > 0$ (or $c_2 < 0$) the particles of $L_A$
belonging to $\Sigma_{t2}$
are driven towards $\Gamma_t$ and necessarily adhere to $S$ along $%
\Sigma_{t1}$. For $u < 0$ (or $c_2 > 0$) the result is reversed:
the particles of $L_A$ belonging to $\Sigma_{t1}$ reach $\Gamma_t$
and are driven towards $ \Sigma_{t2}$.

\begin{figure}[h]
\begin{center}
\includegraphics[width=12.75 cm]{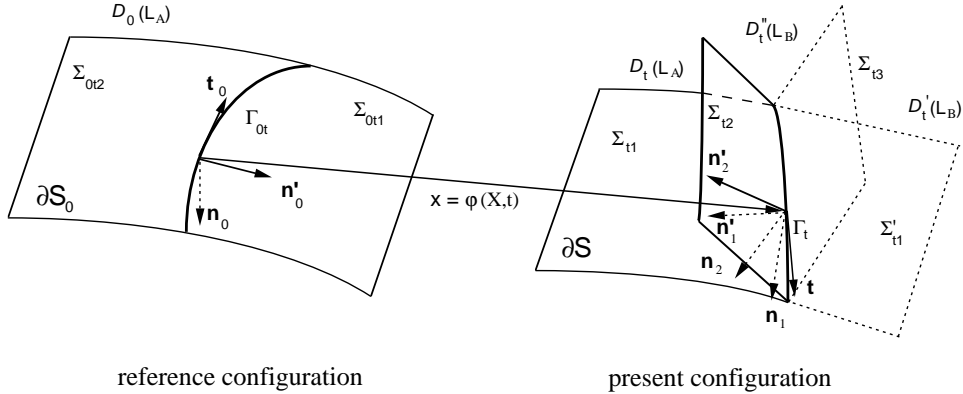}
\end{center}
\caption{In the reference configuration, the two sheets of the
same material surface of fluid $ L_A $ are represented by a
manifold $
\partial S_0$ differentiable along $\Gamma_{0t}$. Its image in  present configuration,
 $D_t$ is divided into two
differentiable manifolds
 $\Sigma_{t1}, \Sigma_{t2}$ forming a dihedral angle. The
common edge $ \Gamma_t$ is the image of a moving curve
$\Gamma_{0t}$ in $\partial S_0$. The triad $\mathbf{n}_0,
\mathbf{t}_0, \mathbf{n}^{\prime}_0,$  in the reference
configuration is transformed by the mapping ${\mbox{{\boldmath
$\varphi$}}} $ into the triad $\mathbf{n}_1, \mathbf{t},
\mathbf{n}^{\prime}_1$ or the triad $\mathbf{n}_2, \mathbf{t},
\mathbf{n}^{\prime}_2$  depending on whether its image is
associated with the manifold $\Sigma _{t1}$ or the manifold
$\Sigma _{t2}$. For other notations see main text.} \label{fig3}
\end{figure}

\noindent In fig. \ref{fig3} the motion of the fluids is sketched.
The two manifolds $\Sigma _{t1}$ and $\Sigma _{t2}$ constitute two
sheets of the same material surface. The motion of the liquid
$L_{A}$ is represented by using a continuous mapping
${\mbox{{\boldmath $\varphi$}}} $ from a half reference space $
D_{0}(L_{A})$ bounded by $\partial S_{0}$, see fig. 3,  onto the
actual domain $ D_{t}(L_{A}) $ occupied by $L_{A}$. The domain
$D_{t}(L_{A})$ is included in the dihedral angle formed by $\Sigma
_{t1}$ and $\Sigma _{t2}$. The contact line $\Gamma _{t}$ is the
image of the mobile curve $\Gamma _{0t}$ on $\partial S_{0}$.
Outside $\Gamma _{0t}$, the mapping ${\mbox{{\boldmath
$\varphi$}}} $ is $ C^{2}$-differentiable.

\noindent A second fluid $ L_B $ occupies the supplemental
dihedral angle $ (\Sigma^{\prime}_{t1},\Sigma_{t2})$. The
conditions of motion are the opposite to those of $L_A$. The
material surface $\Sigma_{t2}$ is the common  interface between
$L_A$ and $L_B$. Provided the fluids are not inviscid, the
velocities of the fluids $L_A$ and $L_B$ are equal along
$\Sigma_{t2}$. Moreover, for liquid $ L_A $, if $ u > 0$, the
particles of $L_A$ are driven towards $\Gamma_t$ and adhere to $S$
along $\Sigma_{t1}$. The particles of $L_B$ are also driven
towards $\Gamma_t$; if they adhere to $S$ along $
\Sigma^{\prime}_{t1}$, the contact line goes through the two
fluids $L_A$ and $L_B$ along the solid wall $\partial S$, (if $u <
0$, a change in the time direction along the trajectories leads to
analogous consequences).
 This is in direct conflict with the fact that  $\Gamma_t$ belongs
 to the interface
 $\Sigma_{t2}$ separating $ L_A $ and $ L_B $.
 To remove this contradiction,
\textit{it is possible to separate} the fluid $L_B$ in two parts
with a material surface $\Sigma_{t3}$ (see figs. \ref{fig2} and
\ref{fig3}). $ \Sigma^{\prime}_{t1}$ and $\Sigma_{t3}$ are the two
sheets of the same material surface for a domain
$D^{\prime}_t(L_B)$ of the fluid $L_B$ within the wedge formed by
the dihedral angle ($\Sigma^{\prime}_{t1}$, $\Sigma_{t3}$). The
sheets $\Sigma_{t2}$ and $\Sigma_{t3}$ constitute two parts of the
same material surface for a domain $D_t^{\prime \prime}(L_B)$ of
the fluid $  L_B $ within the wedge formed by
the dihedral angle ($\Sigma_{t2},\Sigma_{t3}$). The two domains $%
D^{\prime}_t(L_B)$ and $D_t^{\prime \prime}(L_B) $ with common
material surface $ \Sigma_{t3} $ - across which velocity is
continuous - constitute two independent fluid domains which do not
mix. For the domain $D_t^{\prime}(L_B)$, the conditions in the
vicinity of the contact system are similar to those of liquid $
L_A $. Velocities which are discontinuous and multi-valued on
$\Gamma_t$, are compatible with the movements of fluids $ L_A $
and $ L_B $ within the domains $D_t(L_A) $, $D_t^{\prime}(L_B)$ and $%
D_t^{\prime \prime}(L_B)$.

\section{Equations of motion and boundary conditions revisited}

The fundamental law of dynamics is expressed in the form of the
Lagrange-d'Alembert principle of virtual work applied to any
compact domain
 of the two fluids $L_A$ and $L_B$.

\noindent Any compact domain $D_t$ made up of  the two fluids $
L_A $ and $ L_B $ is bounded by $\Sigma_t$. The boundary
$\Sigma_t$ is constituted of $\Sigma_{t1}$, $\Sigma^\prime_{t1}$
and a complementary surface $\Sigma^\prime_{t}$ which is not in
contact with the solid surface $\partial S$: $\Sigma_{t}  =
\Sigma_{t1} \cup \Sigma^\prime_{t1} \cup \Sigma^\prime_{t}$  (fig.
2).
 In a Galilean frame, the virtual work due to the forces applied to
$ L_A $ and $ L_B $\, (including inertial forces but without
forces due to capillarity) is in the general form
$$
 \int_{D_t} [\ ( \phi_i - \rho\ a_i)\ \zeta^i + ( p\ \delta_i^j - Q_i^j)\ \zeta_{,j}^i\ ]\, dv +
\int_{\Sigma_t} P_i \ \zeta^i\, da. \eqno (1)
$$
Here, $dv$, $da$ (and later $dl$) are the volume, area (and later
line) increments, ${\mbox{{\boldmath $\zeta$}}}$ denotes any
virtual displacement field, ${\mbox{{\boldmath $\phi$}}} $ the
volumetric forces, $\rho$ the density, $\mathbf{a} $ the
acceleration vector, $\bf{Q}$ the viscous stress tensor and $p$
the pressure. Moreover, the stress vector  $\mathbf{P}$ describes
the action  of the external media on  $\Sigma_t$. A  contribution
along $\Gamma_t$ is not accounted for.

\medskip

\noindent In continuum mechanics, fluid-fluid and fluid-solid
interfaces are differentiable manifolds endowed with surface
energies\footnote{In statistical physics, fluid interfaces are
transition layers of molecular size. They are modelled in
continuum mechanics with regular surfaces (Rowlinson and Widom,
\cite{row}). On a molecular scale, a solid wall is rough; but
 in
continuum mechanics, when the scale of the roughness is
vanishingly small relative to the size of the solid wall, the
solid wall and the fluid-solid surface energy are modelled with a
{\it differentiable  average  surface, flat on a microscopic
scale} and a {\it corrected surface energy} (Wolansky and Marmur,
\cite{wol}).}. We denote $\sigma _{AB},\sigma _{AS}$ and $\sigma
_{BS}$, the surface energies of interfaces liquid $L_A$-fluid
$L_B$, liquid $L_A$-solid $S$ and fluid $L_B$-solid $S$,
respectively. It is usual to define a measure of energy on
interfaces denoted by $\sigma\, da$, where $\sigma$ stands for
$\sigma _{AB},\sigma _{AS}$ or $\sigma _{BS}$ following the
interfaces between the fluids and the solid. The total energy of
capillarity of the interfaces $\Sigma_{t1}, \Sigma^{\prime}_{t1}$
and $\Sigma_{t2}$ is
\[
E= \int_{\Sigma_t} \sigma \ da.
\]
For any virtual displacement field, the variation of E is (Gouin
and Kosi\'{n}ski, \cite{kosinski}),
\[
\delta E =  \int_{\Sigma_t}\   \Big[\ \delta \sigma -\left(
{\frac{2\sigma }{R_m}}\ n_i +
\left(\delta_i^j-n^jn_i\right)\sigma_{,j}\right)\zeta^i\ \Big]\ da
\]
\[+\int_{\Gamma _{t}}\Big((\sigma _{AS}-\sigma _{BS})\,{n^{\prime
}}_{1i}+\sigma _{AB}\; {n^{\prime }}_{2i}\Big)\ \zeta ^{i}\, dl,
\]
where the scalar $\delta \sigma$ is the variation of the surface
energy $ \sigma$ associated with the displacement
${\mbox{{\boldmath $\zeta$}}}$; vector
 $\mathbf{n} $  ($n^i)$ and scalar $R_m$ stand
  for the   unit normal vector and the mean radius of curvature to $\Sigma_{t1},
 \Sigma^\prime_{t1}$ or $\Sigma_{t2}$,  respectively.\\
The surface energy $\sigma_{AB}$ between the liquid $L_A$ and the
fluid $L_B$ is positive and constant (Rowlinson and Widom,
\cite{row}). Generally, a fluid-solid surface energy depends on
the fluid which is in contact with the solid, the geometrical and
physico-chemical properties of the solid, the microscopic
asperities or the presence of a surfactant. The simplest case
occurs when the surface energy is defined as a function of the
position on the surface ($\mathbf{x} \in \Sigma_{t} \rightarrow
\sigma ( \mathbf{x},t) $).

\noindent Hereafter considering such a case, the virtual work due
to the forces of capillarity applied to $ L_A $ and $ L_B $\, is
simply
$$
  \int_{\Sigma_{t}} {\frac{{2 \sigma} }{R_{m}}}\ n_{i}\ \zeta^i\ da  -
\int_{\Gamma _{t}}\Big((\sigma _{AS}-\sigma _{BS})\,{n^{\prime
}}_{1i}+\sigma _{AB}\; {n^{\prime }}_{2i}\Big)\ \zeta ^{i}\,
dl,\eqno (2)
$$
and relations (1), (2) lead, after execution of the variations and
performing integration by parts in several volume and surface
terms, to the expression, denoted by $\delta T$, of the virtual
work by forces applied to  the domain $D_t$
\[
\delta T=\int_{D_{t}}[\, (\phi _{i}-\rho
a_{i}-p_{,i}+Q_{i,j}^{j})\ \zeta ^{i} \,]\, dv
\]
\[
+\int_{\Sigma^\prime_{t}}\left( P_{i}+p\,n_{i}- Q_{i}^{j}\
n_{j}\right) \zeta ^{i}\ da\, +\int_{\partial S}\left( P_{i}+
({\frac{2{\sigma _{S}}}{R_{m}}}+p)\ n_{1i}- Q_{i}^{j}\
n_{1j}\right) \zeta ^{i}\ da
\]
\[
+\int_{\Sigma _{t2}}\left( (\ {\frac{2{\sigma
_{AB}}}{R_{m}}}+p_{A}-p_{B})\ n_{2i}- (Q_{Ai}^{j}-\ Q_{Bi}^{j})\
n_{2j} \right) \zeta ^{i}\, da
\]
$$
-\int_{\Gamma _{t}}\Big((\sigma _{AS}-\sigma _{BS})\,{n^{\prime
}}_{1i}+\sigma _{AB}\; {n^{\prime }}_{2i}\Big)\ \zeta ^{i}\, dl.
\eqno(3)
$$
 Unit normal vectors
$\mathbf{n_{1}}$ and $\mathbf{n_{2}}$ are exterior to the domain
of liquid $L_A$ and fluid $L_B$, respectively; $\sigma_{S}$ is
called $\sigma _{AS}$ or $\sigma _{BS}$ depending upon which fluid
is in contact with $\partial S$, and $Q_{i}^{j}$ is called
$Q_{Ai}^{j}$ or $Q_{Bi}^{j}$, respectively depending upon which
fluid is in contact with $ \Sigma^\prime_t$ and $\partial S$. We
emphasize that it is not necessary for $D_{t}$,  $\Sigma^\prime
_{t}$ and $\Gamma _{t}$ to be material. Finally, virtual
displacements are tangential to the solid surface $\partial S$ ($
\displaystyle \mathrm{on}\,\,\partial S,\,\,n_{1i}\zeta
^{i}=0$).\newline

\noindent  The expression of the Lagrange-d'Alembert principle is
(Germain, \cite{germain}):
 \begin{center}
 {\it For any } ${\mbox{{\boldmath $\zeta $}}}$  {\it such
 that on}
$\displaystyle \partial S,\,\,n_{1i}\zeta ^{i}=0 ,\ then\ \delta T
= 0$.
 \end{center}

\noindent We emphasize that this principle is not associated with
a variational approach and there is no variational principle in
it: $\delta T$ is not the Frechet derivative of a functional
(Gurtin, \cite{gurtin}). Only for equilibrium, and due to the fact
that the viscous stress tensor is null, the minimization of energy
(this is a variational principle) coincides with this approach.
Such a method is relevant to the theory of distributions where
${\mbox{{\boldmath $\zeta $}}}$ are vector fields of class
$C^{\infty}$ with compact support  (Schwartz, \cite{schwartz}).

\noindent The equations of motion and natural boundary conditions
that emerge from it are as follows:
\\

{\bf Equations of motion}
$$
\rho a_i + p_{,i} = \phi_i + Q^j_{i,j}. \eqno (4)
$$

{\bf Conditions on the liquid $L_A$ - fluid $L_B$ interface}
$\Sigma_{t2}$

$$
{\frac{{2\sigma _{AB}}}{R_{m}}} n_{2i}=(Q_{Ai}^{j}-Q_{Bi}^{j})
n_{2j}+(p_{B}-p_{A})n_{2i}, \eqno(5)
$$
which is the dynamic form of the Laplace equation.
\\

{\bf Conditions on the boundary} $\Sigma^\prime_t$

$$
P_{i} = - p\,n_{i}+ Q_{i}^{j}\, n_{j}, \eqno(6)
$$
which is the classical expression of the balance of stresses for
viscous fluids.
\\

{\bf Conditions on the surface} $\partial S =  \Sigma_{t1} \cup
\Sigma^\prime_{t1}
$\\

\noindent Expression (3) of the virtual work and the
Lagrange-d'Alembert principle imply: For any  ${\mbox{{\boldmath
$\zeta $}}}$  such that $\displaystyle \mathrm{on}\,\,\partial
S,\,n_{1i}\zeta ^{i}=0,$
\[
\int_{\partial
S}\left( P_{i}+ ({\frac{2{\sigma _{S}}}{R_{m}}}+p)\ n_{1i}-
Q_{i}^{j}\ n_{1j}\right) \zeta ^{i}\ da = 0.
\]
Consequently, there exists
 a scalar field $\chi $ of Lagrange
multipliers defined on $\partial S$ such that  (Kolmogorov and
Fomin, \cite{kolmogorov})
$$
P_{i}\ = Q_{i}^{j}\, n_{1j}- ({\frac{{2\sigma _{S}}}{R_{m}}} + p)\
n_{1i}+ \chi \ n_{1i}. \eqno(7)
$$
Generally $\mathbf{P}$ is not collinear to $\mathbf{n}_{1}$ and
$\chi $ is an additional unknown scalar.
\\

{\bf  Conditions on the contact line} $\Gamma_t$\\

\begin{figure}[h]
\begin{center}
\includegraphics[width=7.5 cm]{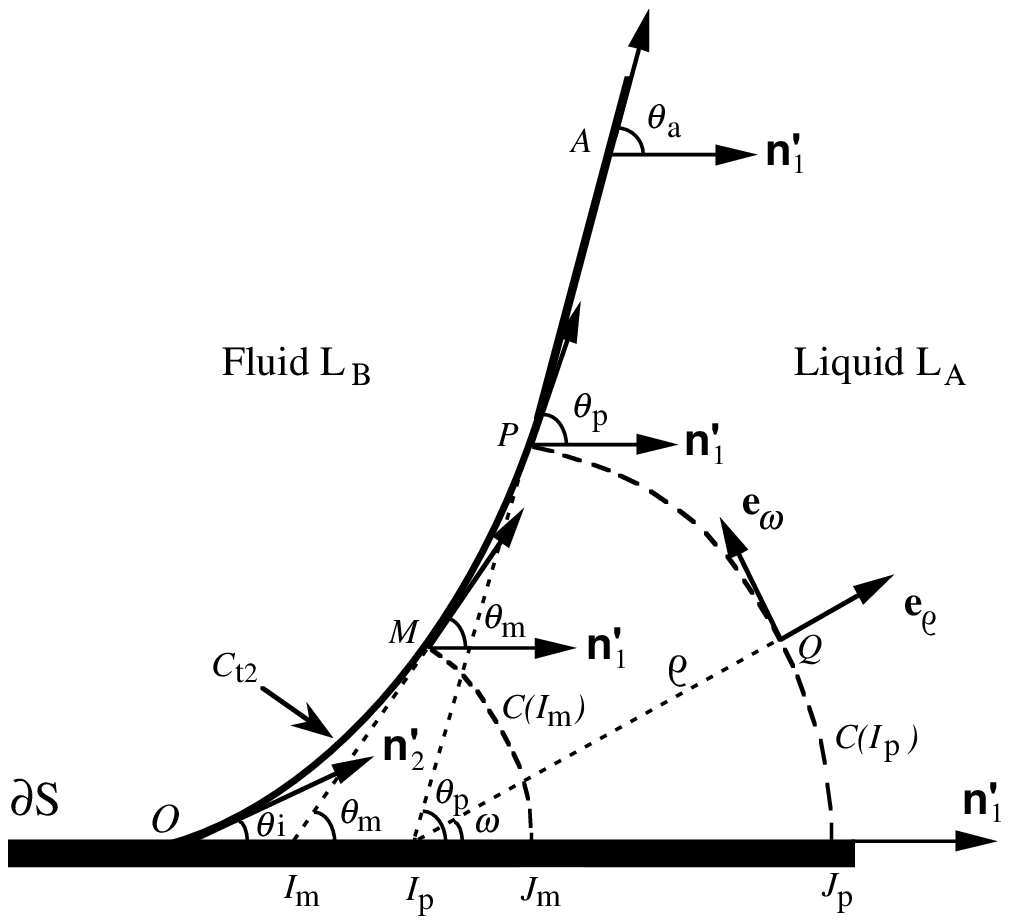}
\end{center}
\caption{Cross section $C_{t2}$ of the fluid-fluid interface and
the solid wall in the plane $(O,\mathbf{n^{\prime
}}_{1},\mathbf{n^{\prime }}_{2})$. The contact line is reduced in
this figure to the point $O$. The intrinsic angle $\theta_i$ is
the angle between $\mathbf{n^{\prime }}_{1}$ and
$\mathbf{n^{\prime }}_{2}$. At any generic point of $C_{t2}$, the
angle between $\mathbf{n^{\prime }}_{1}$  and the tangent to
$C_{t2}$ is denoted by $\theta$; thus, at the points $M$, $P$  and
$A$, the angle $\theta$ is denoted  by $\theta_m$, $\theta_p$ and
$\theta_a$, (see main text in section 3). The intersections of the
tangent lines to $C_{t2}$ in $M$ and $P$ with the axis $ O\,
\mathbf{n^{\prime }}_{1}$  are denoted by $I_m$ and $I_p$. The
arcs of circles $C(I_m)$ and $C(I_p)$  of centers $I_m$ and $I_p$
and radius $I_mM$ and $I_pP$ intersect the axis $ O\,
\mathbf{n^{\prime }}_{1}$ at points $J_m$ and $J_p$, respectively.
A point $Q$ of $C(I_p)$ is represented by   $(\varrho,\omega)$ in
the polar coordinate system of pole $I_p$ and polar axis $ O\,
\mathbf{n^{\prime }}_{1}$ (see main text in subsection 5.2). }
\label{fig4}
\end{figure}

\noindent Due to the condition  on $\partial S,\ n_{1i}\zeta
^{i}=0$, a virtual displacement is expressed at any point of the
contact line $ \Gamma_t$ in the form
$$
{\mbox{{\boldmath $\zeta$}}} =\kappa\ \mathbf{t} + \upsilon\
\mathbf{ n^{\prime} }_1, \eqno (8)
$$
\noindent where the two scalar fields $\kappa $ and $\upsilon $
are defined on $ \Gamma _{t}$. For any field ${\mbox{{\boldmath
$\zeta$}}}$ in the form (8), the contribution of $\, \int_{\Gamma
_{t}}\Big( (\sigma _{AS}-\sigma _{BS})\;n_{1i}^{\prime }+\sigma
_{AB}\;n_{2i}^{\prime }\Big) \,\zeta ^{i}\, dl  $ in relation (3)
yields
$$
\int_{\Gamma _{t}}\Big( (\sigma _{AS}-\sigma
_{BS})\;n_{1i}^{\prime }+\sigma _{AB}\;n_{2i}^{\prime }\Big)
\,(\kappa\ t^i + \upsilon\ {{n^{\prime }}_1}^i )\, dl = 0 \eqno(9)
$$
In the general case, since $\mathbf{n^{\prime }}_{1}=
\mathbf{n}_{1} \times \mathbf{t} $ and  $\mathbf{n^{\prime }}_{2}=
\mathbf{n}_{2} \times \mathbf{t} $, expression (9) implies
$$
\sigma _{AS}-\sigma _{BS}\ +\sigma _{AB} \, \mathrm{cos}\ \theta
_{i}\,=0 ,\eqno (10)
$$
where $\theta _{i}$ is the angle between $\mathbf{n^{\prime
}}_{1}$ and $\mathbf{n^{\prime }}_{2}$. This angle named
\textit{intrinsic contact angle} in the literature  (Wolansky and
Marmur, \cite{wol}), is the angle in a plane that is normal to
$\partial S$ and $\Gamma_t$ between tangents in $O$ parallel to
$\Sigma _{t2}$ and $\partial S$  (see fig. \ref{fig4}).

\noindent At point $O$ of the contact line $\Gamma _{t},$ let us
consider the section of the $L_A$-$L_B$ interface in the plane
erected by $\ \mathbf{n^{\prime }}_{1}$ and $\mathbf{n^{\prime
}}_{2}$. Let the curvature of the planar section $C_{t2}$ of
$\Sigma _{t2}$ be $R^{-1}$. For  a two-dimensional flow, the mean
curvature of the surface $\Sigma _{t2}$ is $ 2\, R_m^{-1} =
R^{-1}$. At a generic point of $C_{t2}$, the angle between
$\mathbf{n^{\prime }}_{1}$ and the tangent to $C_{t2}$  is denoted
by $\theta$.   This angle depends on the choice of the point and
on the fluid flow. Since the surface energy $\sigma _{AB}$ between
two fluids is constant and $ R = dl/d\theta$, we obtain,
$$
\int_{O}^{P}\frac{\sigma _{AB}}{R}\mathrm{\ }\sin \theta \ dl
\equiv \sigma _{AB}(\cos \theta _{i}-\cos \theta _{p}) , \eqno(11)
$$
where $\theta_p$ is the value of $\theta$ at the point $P$.
Relation (5) yields
\[
\displaystyle \frac{ \sigma _{AB}}{R } =
(Q_{Ai}^{j}-Q_{Bi}^{j})n_{2}^{i}n_{2j}+ p_{B}-p_{A},
\]
and, consequently, relations (10), (11) yield \footnote{ Relation
(12) can be proved directly by using the projection on $\mathbf{
n^{\prime }}_{1}$ of the balance of forces applied to a liquid
$L_A$ - fluid $L_B$ domain containing the fluid interface $\Sigma
_{t2}$.}
$$
\sigma _{AS}-\sigma _{BS} +\sigma _{AB}\, \mathrm{cos}\ \theta
_{p}\,+\int_{O}^{P}\left( (Q_{Ai}^{j}-Q_{Bi}^{j})\
n_{2}^{i}n_{2j}+p_B-p_A\right) \mathrm{sin}\ \theta \ dl=0.
\eqno(12)
$$

\section{A creeping flow example of non-Newtonian fluids}

\subsection{The Huh and Scriven model revisited}

To understand more precisely the behaviour of a liquid near a
moving solid-liquid-fluid contact line, we reconsider the
situation of two-dimensional flows proposed by Huh and Scriven,
\cite{huh}. Let us recall the main results of their article (see
fig. \ref{fig2}):

\noindent A flat solid surface in translation at a steady velocity
$\mathbf{U}$  is inclined  from a flat interface between a liquid
$L_A$ and a fluid $L_B$ (here, the angle $\theta$ of fig. 4 is
constant, independent of the generic point of $C_{t2}$). The
contact line velocity with respect to the solid is $-\mathbf{U}$;
thus in the notations of section 2, we obtain $ \mathbf{U} = u\,
\mathbf{n}_{1}^{\prime }$. In a two-dimensional situation of the
plane $(O, \mathbf{n}_{1}^{\prime }, \mathbf{n}_{2}^{\prime })$,
it is convenient to take the contact line intersection point $O$
as the origin of a polar coordinate system $(r,\varphi )$ and $O
\, \mathbf{n}_{1}^{\prime }$ as the reference polar axis. The two
bulks are incompressible Newtonian fluids. In term of the stream
function $ \psi (r,\varphi )$ for two-dimensional steady flows,
the velocity is
\[
\mathbf{V}=\;v_{r}\mathbf{e}_{r}+v_{\varphi }\mathbf{e}_{\varphi
}, \, {\rm with}\,\,\,  v_{r}=-{1\over r} {\partial \psi
\over\partial \varphi}, \, v_{\varphi }= {\partial \psi
\over\partial r},
\]
 and ($O, \mathbf{e}%
_{r}, \mathbf{e}_{\varphi })$ as the mobile polar frame. In the
creeping flow approximation of a viscous fluid,  Eq. (4) leads to
the linearized Navier-Stokes equation and consequently to the
biharmonic equation $\displaystyle\nabla ^{4}\psi =0\,$, where
$\displaystyle\nabla ^{2}$  and $\displaystyle\nabla ^{4}$
 are respectively the Laplacian and bi-Laplacian operators
 (Moffat, \cite {moffat}; Bataille, \cite{bat}). A solution is in
the form:
\[
\psi (r,\varphi )\equiv rf(\varphi )
\]
which leads to the ordinary differential equation
\[
f(\varphi)+2f^{\prime \prime} (\varphi)+f^{(IV)}(\varphi)=0,
\]
and thus to the general solution
$$
\psi (r,\varphi )=r (a\ \mathrm{sin}\ \varphi +b\ \mathrm{cos}\
\varphi +c\ \varphi \ \mathrm{sin}\ \varphi +d\ \varphi \
\mathrm{cos}\ \varphi ),\eqno (13)_1
$$
which holds for either fluid.

\noindent The boundary conditions at the solid wall and the
liquid-fluid interface are:

$(i)$ a vanishing normal component of the velocity at the solid
surface and interface,

$(ii)$ continuity of the velocity at the interface,

$(iii)$ continuity of the tangential stress at the interface,

$(iv)$ non tangential relative motion of the fluids at the solid
surface except at the contact line.

\noindent These eight linear conditions yield the values of
coefficients $a, b, c, d$ for the two fluids $L_A$ and $L_B$. If
the dynamic viscosity coefficients are identical, the eight
integration constants are:
\[
\left\{
\begin{array}{l}
\displaystyle a_A = -u\, \theta D(\theta) [\ \pi - \theta
+\mathrm{sin}\
\theta \ \mathrm{cos}\ \theta \ ], \\
\displaystyle a_B = u\,D(\theta) [\ (\pi-\theta ) (\mathrm{sin}\
\theta \
\mathrm{cos}\ \theta - \theta )+\pi \ \theta \ \mathrm{sin}^2\ \theta\ ], \\
\displaystyle b_A = 0, \\
\displaystyle b_B = u\, D(\theta) [\ \pi \ \theta\ \mathrm{sin}\
\theta \
\mathrm{cos}\ \theta - \pi \ \mathrm{sin}^2\ \theta\ ],  \hskip 4.5 cm (13)_2\\
\displaystyle c_A = u\, D(\theta)\ (\pi - \theta ) \ \mathrm{sin}^2\ \theta, \\
\displaystyle c_B = - u\, D(\theta) \ \theta \ \mathrm{sin}^2\ \theta, \\
\displaystyle d_A = u\, D(\theta) [\ (\pi -\theta)\ \mathrm{sin} \
\theta \
\mathrm{cos}\ \theta + \ \mathrm{sin}^2\ \theta\ ], \\
\displaystyle d_B = u\, D(\theta) [\ -\theta\ \mathrm{sin}\ \theta
\ \mathrm{ cos}\ \theta + \ \mathrm{sin}^2\ \theta\ ]\

\end{array}
\right.
\]
with $D(\theta ) =[\ \theta \ (\pi -\theta ) - (\pi -2 \theta)\ \mathrm{sin}%
\ \theta\ \mathrm{cos}\ \theta - \mathrm{sin}^2\ \theta \ ]^{-1}$.

\noindent No difficulty should arise, in principle, in the
determination of $a_A,\ldots, d_B$ for two fluids with distinct
dynamic viscosity coefficients. The form of the streamlines as
obtained by Huh and Scriven are sketched in fig. \ref{fig2}, and
the motion of the contact line fits perfectly with the kinematics
as outlined in section 2. Furthermore, if $\mu _{A}=\mu _{B}=\mu $
for all  values of the dynamic contact angle $\theta$, the viscous
stress components are
\[
\displaystyle\tau _{r\varphi }=-\;\frac{\mu }{r}(f+f^{\prime
\prime})\ \equiv \ -\ 2\ \frac{ \mu }{r}(c\ \mathrm{cos}\ \varphi
-d\ \mathrm{sin}\ \varphi ),\  \tau _{rr}=\tau _{\varphi \varphi
}=0
\]
and the pressure field is given by
\[
p=p_{0}+\frac{\mu }{r}\;(f^{\prime}+f^{\prime \prime
\prime})\equiv p_{0}-2\ \frac{\mu }{r}(c\ \mathrm{sin}\ \varphi \
+d\ \mathrm{cos}\ \varphi ),
\]
where $p_{0}$ is the hydrostatic pressure (in both formulae phase subscripts
have been omitted).

\noindent As proved by Huh and Scriven, the dissipation in any
domain $D_t$ of the fluids containing the contact line $\Gamma_t$,
  $  \Big ( \int_{D_{t}} (\mu/2)\, (f+f^{\prime \prime})^{2}/{r^2}\  dv  \Big ),
$ and the total traction exerted on the solid surface by the fluid
interface are logarithmically infinite. Moreover the normal stress
across the fluid interface varies as $r^{-1}$; furthermore, the
stress jump should be balanced by the Laplace interfacial tension
$ \displaystyle \sigma _{AB}\,R^{-1}$ and the curvature $R^{-1}$
does increase indefinitely at the contact line. These are all
non-integrable singularities.

\subsection{A model of non-Newtonian fluid near the contact line}

To avoid the previous paradox of an infinite dissipative function
at the contact line, we consider non-Newtonian incompressible
fluids with a convenient behaviour of the viscous stress tensor.
It is experimentally known that the dynamic viscosity $\mu $ of
polymeric liquids depends on the shear rate $\dot{\epsilon}$. The
behaviour prevailing in such situations is not well understood. In
a wide variety of technological applications, liquids are
subjected to large shear strain forces. A molecular dynamic
investigation of liquids subjected to large shear strain rates has
been performed by Heyes {\it et al}, \cite{heyes}. The shearing
action has been found to change the liquid structure and reveals a
tendency to reduce the shear viscosity (Ryckaert {\it et al},
\cite{ryckaert}).\newline In the literature, some empirical
formulas for the viscosity obtained by means of a weighted
least-squared adjustment were proposed. For example, $\mu
(\dot{\epsilon} )\simeq \;\mu (0)\;-\;c$ $\dot{\epsilon}^{2}$ was
suggested as possible form for the viscosity as a function of
$\dot{\epsilon} $ (Heyes {\it et al}   \cite{heyes}). Holian and
Evans, \cite{Evans}, proposed a representation in the form
$$
\mu (\dot{\epsilon} )\simeq \;\mu
(0)\;-\;c\sqrt{\dot{\epsilon}}.\eqno (14)
$$
Data for the viscosity of an atomic fluid generated by
nonequilibrium molecular dynamic were performed by Ryckaert {\it
et al}, \cite{ryckaert}. They indicated that for shear rates below
$10^{12}s^{-1}$, $\mu (\dot{\epsilon} )$ does not differ
significantly from $\mu (0)$ but that these previous laws are not
extendable when $\dot{\epsilon}$ tends to infinity.

\noindent We propose a model where the viscous stress tensor
$\mathbf{Q}\;$is a function of the  strain rate tensor
$\displaystyle{\mbox{{\boldmath $\Delta$}}} \equiv ({1/2})\left(
\nabla \mathbf{V}+(\nabla \mathbf{V})^{t}\right) $. For moderate
values of ${\mbox{{\boldmath $\Delta $}}} $ the fluid is Newtonian
and the function is linear. The function deviates from this
classical behaviour for large values of ${\mbox{{\boldmath $\Delta
$}}} $. For an isotropic stress tensor of {\it two-dimensional}
flow, the Rivlin-Ericksen representation theorem (Truesdell and
Noll, \cite{truesdell}) leads to a viscous stress tensor in the
form $\mathbf{Q}=\lambda \ \mathbf{I}+2\ \mu {\mbox{{\boldmath
$\Delta$}}} $ but $ \lambda $ and $\mu $ are non-constant
functions of invariants of ${\mbox{{\boldmath $\Delta $}}}, $ and
$\mu {\mbox{{\boldmath $\Delta $}}} $ is a functional of
${\mbox{{\boldmath $\Delta $}}} $ where $\mu $ tends to $\mu _{0}\
$($\mu _{0}$  being constant) when ${\mbox{{\boldmath $\Delta $}}}
$ tends to zero.
\newline We propose to use a convenient representation of the $\mu
$-behaviour in the form
$$
\mu =(1-e^{-\gamma })\ \mu _{0}\,\,\,\,\mathrm{with}\,\,\,\,\gamma
=\left( {     \frac{1}{\left\| {\mbox{{\boldmath $\Delta $}}}
\right\| \tau _{0}}}\right) ^{\xi }\eqno(15)
$$
where $\left\| {\mbox{{\boldmath $\Delta $}}} \right\| \;=\;\tau
^{-1}\;$is the norm of the strain rate tensor, $\tau _{0}$ is a
characteristic time of the fluid and $\xi $ is a small parameter
($0<\xi \ll 1$). Then, for very high shear rates $\mu $ behaves as
a step function as expected in Ryckaert {\it et al}. To fit with
relation (14) when $\dot{\epsilon}$ is close to $10^{12}s^{-1},$
we choose $ \xi \;=\;0.2\;$ and $\tau _{0}=10^{-14}s$ but many
other values can be considered  and results of the literature are
disparate.
\newline In the following, we take  $\xi = 0.1\;$and
$\tau _{0}=10^{-12}s $ ; then for $\gamma =\gamma_0 \equiv 4.6 $,
we obtain $\mu \;=0.99\,\mu _{0}$. We call this $\gamma
_{0}$-value the \textit{cut-off} coefficient. For $\gamma =\gamma
_{0},\; \mu \simeq \mu _{0},\;$and the fluid may be considered as
Newtonian.\newline
\newline
For the Huh and Scriven model  of two-dimensional incompressible
flows, $ \left\| {\mbox{{\boldmath $\Delta $}}} \right\| =
\dot{\epsilon} =\displaystyle  \left| f+f^{\prime \prime}\right|
/(2\ r)$.$\;$ When $u=1\ mm.\ s^{-1}$, due to the $\gamma
_{0}-$value, considerable variations of $\mu $ occur from the
contact line to a distance of  $20$ to $30\;$Angstr\"{o}ms. The
same holds true on the solid wall when $\theta \in
\left[ 5^{%
{{}^\circ}%
},175^{%
{{}^\circ}%
}\right] $. Outside these distances from the contact line, the
fluids can be considered as Newtonian. Then, $\mu $ tends to zero
for very large values of the shear rate, and ${\bf Q}$ is a
function of ${\mbox{{\boldmath $\Delta $}}} $ which tends to
infinity with ${\mbox{{\boldmath $\Delta $}}}, $ but weaker than a
linear function.
 The total stress tensor of a fluid is always of the form
$-\;p\;$\textbf{Id} $+\;2\;\mu {\mbox{{\boldmath $\Delta $}}} \;$
where $p\;=\Pi-\lambda$ (here $\Pi $ notes the hydrostatic
pressure).
 For steady flows, the
equation of motion is
$$
\rho\ \mathrm{grad}\ ({\frac{1}{2}}\mathbf{V}^{2})+\rho\
\mathrm{rot}\ \mathbf{V}\times \mathbf{V}+\mathrm{grad}\ p=\mu\,
\nabla ^{2}\mathbf{V}+2\ {\mbox{{\boldmath $\Delta $}}}
{\boldmath.} \mathrm{grad}\ \mu +\rho\ \mathbf{g}\eqno(16)
$$
where $\mathbf{g}$ denotes the acceleration due to gravity.
 For a stream function $\psi (r,\varphi )\equiv
rf(\varphi )$, we obtain
\[
\mu \nabla ^{2}\mathbf{V}=-\frac{\mu_0\ (1-e^{-\gamma})} {r^{2}
}\left( ( f^{\prime }+f^{\prime \prime \prime} )
\mathbf{e}_{r}+(f+f^{\prime \prime})\mathbf{e}_{\varphi }\right),
\]
\[
2\ {\mbox{{\boldmath $\Delta $}}} {\boldmath.} \mathrm{grad}\ \mu=\xi \ \mu _{0}\ \frac{\gamma \ e^{-\gamma } }{%
r^{2}}\ \left( ( f^{\prime }+f^{\prime \prime \prime} ) \mathbf{e}_{r}-(f+f^{\prime \prime}) \mathbf{e}%
_{\varphi }\right),
\]
\[
\rho\ \mathrm{rot}\ \mathbf{V}\times
\mathbf{V\;=}\;-\frac{\rho}{r} \;(f+f^{\prime \prime})\left(
f\mathbf{e}_{r}+f'\mathbf{e}_{\varphi }\right).
\]
The inequalities $0\;<\;\gamma \ e^{-\gamma }\;\leq\;1-e^{-\gamma }$, $%
\;0<\;\xi \ll \;1$ \ and the fact that near the contact line $\
1\ll \;\displaystyle \frac{\mu}{r \rho\ (f+f^{\prime\prime})}\ $
yields the approximate form of Eq. (16)
\[
\rho \ \mathrm{grad}\ ({\frac{1}{2}}\mathbf{V}^{2})+\mathrm{grad}\
p =\mu \nabla ^{2}\mathbf{V}+\rho\ \mathbf{g},
\]
which implies
$$
\mu \mathrm{\ rot(}\nabla ^{2}\mathbf{V)}+ \mathrm{grad\;\mu \;\times \;} \nabla
^{2}\mathbf{V\;=0}\eqno(17)
$$
with
\[
\mu \ \mathrm{\ rot(}\nabla ^{2}\mathbf{V)\;=-\;}\frac{\mu
_{0}}{r^{3}}\ (1-e^{-\gamma })\
(f+2f^{\prime\prime}+f^{(IV)})\;\mathbf{k}
\]
and
\[
\mathrm{grad\;\mu \;\times \;}\nabla ^{2}\mathbf{V\;=\;}\xi \ \mu
_{0}\ \frac{\gamma \ e^{-\gamma }}{r^{3}}\left(
f+f^{\prime\prime}+{\frac{(f^{\prime }+f^{\prime \prime
\prime})^{2}}{f+f^{\prime\prime}}}\right) \mathbf{k}
\]
where $\mathbf{k}$ denotes the normal vector to the plane of the
flows.

\noindent From $\ \xi \ \gamma \ e^{-\gamma }\ll \;1-e^{-\gamma
},$ we deduce again the Huh and Scriven approximation for the
stream function in the form $\psi (r,\varphi) = r f(\varphi)$,
with $ f(\varphi)+2f^{\prime
\prime}(\varphi)+f^{(IV)}(\varphi)=0$.\newline
\newline
On the solid wall, adherence conditions are required. This
assumption is in agreement with molecular dynamics of fluid flows
at solid surfaces: \textit{\ The non-slip boundary condition
appears to be a natural property of a dense liquid interacting
with a solid wall with molecular structure and long range force
interactions} (Koplick, Banavar and Willemsen, \cite{kop}). For
the non-Newtonian model and for the creeping flow approximation,
the general solution $(13)_1$  holds true for either fluid.
Furthermore, the boundary conditions at the solid wall and at the
liquid-fluid interface are the condition $(i)-(iv)$ of subsection
4.1. Consequently, for the non-Newtonian model and for the
creeping flow approximation the trajectories and the velocities
near the contact line are identical to those of fluids with
constant viscosity $\mu _{0}$ in the Huh and Scriven model.
\newline
\newline
The dissipation $\Xi $ in the domain $V_{l}=\{r\in \lbrack 0,l],\
\varphi \in \lbrack 0,\theta ],\ z\in \lbrack 0,L]\}$ (where $z$
denotes the contact line coordinate) is
\[
\Xi \ =\int_{V_{l}}tr(\mu {\mbox{{\boldmath $\Delta $}}}
^{2})dv=L\ {\frac{\mu _{0}}{2}}\int_{0}^{\theta
}\int_{0}^{l}{\frac{{\ 1-e^{-\gamma }}}{r}}\ (f+f^{\prime
\prime})^{2}\ dr\ d\varphi.
\]
From the inequality $\ 0<1-e^{-\gamma }\leq \gamma $, we deduce
\[
\Xi \ \leq \ {\frac{L\ \mu _{0}}{2^{1-\xi }}}\int_{0}^{\theta }\int_{0}^{l}{%
\ \ \frac{r^{\xi -1}}{\tau _{0}^{\xi }}}\ |f+f^{\prime \prime}|^{2-\xi }\ drd\varphi ={%
\frac{{\ \ L\ \mu _{0\;}l^{\xi }}}{{\xi \ 2^{1-\xi }\ \tau
_{0}^{\xi }}}} \int_{0}^{\theta }|f+f^{\prime \prime}|^{2-\xi }\
d\varphi,
\]
which proves that the dissipation is finite at the contact line,
since $f$ given by $(13)_1, (13)_2$ is bounded.

\noindent Other calculations yield a bounded total force exerted
on the solid surface by the fluid-fluid interface near the contact
line, but another problem associated with the fluid-fluid
interface curvature still remains unresolved.

\section{Study of a curved   interface in the vicinity of the contact line}

For a Newtonian fluid in two dimensional flows, the fluid
interface curvature should increase rapidly as the contact line is
approached. This result is in direct conflict with the hypothesis
of section 4 that the fluid interface is perfectly flat. Indeed,
Huh and Scriven, \cite{huh}, pointed out \textit{when water at
moderate dynamic contact angle wets a surface at $6$
mm.min$^{-1}$, the local radius of curvature would have to be
about $10^{5}$ time greater than the distance to the contact line
and the curvature would be imperceptible by optical means}.
Nevertheless in such a case, the intrinsic angle $\theta_i$ at the
contact line may strongly deviate from the angle $\theta_p$, (see
fig. 4), which is observed at a point $P$   near, but not at, the
contact line.

\noindent Let us consider results presented in subsection 4.2; the
equation of motion  (4), boundary condition  (5), and calculations
of subsection 4.2 yield the pressure field values for the fluids $
L_{A} $ and $  L_{B} $,
$$
p_{B}-p_{A}\;=\;\frac{\mu }{r}\left( (f_{B}^{\prime
}+f_{B}^{\prime \prime \prime})-(f_{A}^{\prime }+f_{A}^{\prime
\prime \prime})\right) \equiv \frac{ \,\mu\, u}{r}\ 2  \pi\,
\mathrm{sin}\ \theta\ D  (\theta ).  \eqno(18)
$$
As done by Huh and Scriven, we notice that
\[
p_{B}-p_{A}\ =\;\frac{\sigma _{AB}}{R}   , \]
and consequently,
$$ \frac{r}{R}=
\frac{\mu \;u}{\sigma _{AB}}\ 2\pi\, \mathrm{sin\;} \theta\
D(\theta ). \eqno(19)
$$
For partial wetting, it is easy to compute  the value of $2\pi\
\mathrm{sin\,}\theta\ D(\theta ) $ numerically; we obtain
$$
\frac{\pi}{6} < \theta <\frac{5\;\pi}{6}\ \ \Rightarrow \ \   4
<2\pi\, \mathrm{sin\,}\theta\ D(\theta )< 15 .
$$
When the capillary number $C_{a}=  \mu _{0}\,|u|  / \sigma _{AB}\
$is sufficiently small,
 (in experiments,  $C_{a}$ is often smaller than $10^{-3}$), we deduce
 $\displaystyle |r/R|
\ll 1$;   it is all the more true, for a non-Newtonian fluid given
by the representation (15), where $\mu $ tends to zero at the
contact line.

\subsection{ Two-dimensional steady flows near the contact line}
The conditions and the notations are given in subsection 4.2, but
the fluid-fluid interface is curved. The cross section of the
fluid-fluid interface and the solid wall is presented on fig. 4.
We consider the domain occupied by the two fluids in the immediate
vicinity of the contact line and we assume that, along $ C_{t2} $
$$
\lim_{r \rightarrow \,0\,\,}\frac{r}{R} \equiv  \lim_{r
\rightarrow \,0\,\,}\frac{r\, d\theta}{dl}= 0. \eqno (20)
$$
Thus, in the immediate vicinity of the contact line, we obtain
$|r/R| \ll 1$.
\newline On fig. 4, at the generic point $P$ of $ C_{t2} $, the
intersection of the tangent line with the axis $0\, {\mathbf
n}^{\prime}_1$ is denoted by $I_p$. To each point $I_p$
corresponds, in the fluid domains, the arc of a circle, denoted by
$C(I_p)$, with center $I_p$ and radius $I_p P$. To a  point $Q$ of
$C(I_p)$ corresponds the polar coordinates, $(\varrho,\omega)$,
associated with the pole $I_p$ and the mobile frame $({\mathbf
e}_\varrho, {\mathbf e}_\omega)$. The polar coordinates of $P$ are
$(\varrho,\theta_p)$ and will be denoted simply by
$(\varrho,\theta)$. Let us denote by $y \equiv \varrho\, {\rm
sin}\, \theta$, the distance from the point $P$ to the solid wall.
We deduce, $dy = d\varrho\, {\rm sin}\, \theta + \varrho\, {\rm
cos}\, \theta\, d\theta$ and due to the differential relation, $dy
= dl\, {\rm sin}\, \theta $, we obtain, $d\varrho/dl = 1 -\varrho\
{\rm cotg}\, \theta \ d\theta/dl$. When $\theta\neq 0\ {\rm or}\
\pi ,\ \lim_{r \rightarrow \,0\,\,}\varrho\ {\rm cotg}\, \theta \,
d\theta/dl = 0$ and consequently, when $r \rightarrow 0 $,
$d\varrho\sim dl$. Let $M$ be another point of $ C_{t2} $; we
denote $J_p$ and $J_m$ the intersections of $C(I_p)$ and $C(I_m)$
with the axis $O\, {\mathbf n}^{\prime}_1$; when $M\rightarrow P$,
$\|{\mathbf J}_m{\mathbf J}_p\| \sim d\varrho$, $\|{\mathbf
I}_m{\mathbf I}_p\| \sim \varrho\ d\theta/ {\rm sin}\, \theta $
and relation (20) implies, $\lim_{r \rightarrow \,0} (\|{\mathbf
I}_m{\mathbf I}_p\|/\|{\mathbf J}_m{\mathbf J}_p\|) = 0$. Thus, in
the immediate vicinity of the point $O$, $\|{\mathbf I}_m{\mathbf
I}_p\| \ll \|{\mathbf J}_m{\mathbf J}_p\|$ and the two arcs of
circles $C(I_m)$ and $C(I_p)$ are distinct. In the following, we
prove that, {\it near the point $O$, a point $Q$ of the fluid
domains
 is represented by the
orthogonal coordinate system} $(\varrho, \omega)$. The equation of
the curve $ C_{t2} $ can be written in the form $\theta =
\vartheta(\varrho)$.

\noindent As in section 4   for two-dimensional steady flow,  the
stream function $\Psi(\varrho,\omega)$ verifies $\nabla^4 \Psi =
0$. We look for a stream function in the form
\[ \Psi(\varrho,\omega) \equiv  \varrho\ h(\omega,\theta),\]
where $\theta = \vartheta(\varrho)$, and such that the partial
derivatives of $h$ with respect to $\omega$ and $\theta$ are
bounded. But,
\[ d{\mathbf {OQ}} =  d{\mathbf {OI}_p} + d{{\bf I}_p{\bf Q} } =
\left (\, \frac{\varrho\ d\theta}{{\rm sin}\, \theta \ d\varrho}\
{\mathbf n}^{\prime}_1 + {\mathbf e}_\varrho\right)   d\varrho +
\varrho \ d\omega\ {\mathbf e}_\omega.\] Since $ \lim_{\varrho
\rightarrow \,0}\ \varrho\ d\theta/d\varrho = 0$ and $\sin\ \theta
\neq 0$, we obtain
\[ d{\mathbf {OQ}} = d\varrho\ ({\mathbf e}_\varrho + {\mathbf
o}_1 (\varrho)) + \varrho\ d\omega\ {\mathbf e}_\omega, \ \ {\rm
with} \ \ \lim_{\varrho \rightarrow \,0}\ {\mathbf o}_1(\varrho) =
{\mathbf 0},
\]
and near the point $O$
\[ d{\mathbf {OQ}} = d\varrho\  {\mathbf e}_\varrho  + \varrho\ d\omega\ {\mathbf
e}_\omega.
\]
Moreover,
\[ {\rm grad}\ \Psi = ( h +
\varrho\ \frac{\partial h}{\partial \theta}\
\frac{d\theta}{d\varrho} )\ {\mathbf e}_\varrho + \frac{\partial
h}{\partial \omega}\ {\mathbf e}_\omega .\] and
\[ {\rm grad}\ \Psi = ( h + o_2(\varrho)
  )\ {\mathbf e}_\varrho + \frac{\partial
h}{\partial \omega}\ {\mathbf e}_\omega , \ \  {\rm with}\ \
\lim_{\varrho \rightarrow \,0}\  o_2(\varrho) = 0.
\]
The velocity is
\[{\mathbf V} = v_\varrho\  {\mathbf e}_\varrho + v_\omega\ {\mathbf
e}_\omega +\, {\mathbf o}_3 (\varrho)   \ \  {\rm with},\ \
v_\varrho = - \frac{\partial h}{\partial\omega},\ v_\omega = h \ \
{\rm and},\ \ \lim_{\varrho \rightarrow \,0}\ {\mathbf
o}_3(\varrho) = {\mathbf 0},
\]
and near the contact line
\[{\mathbf V} = v_\varrho\  {\mathbf e}_\varrho + v_\omega\ {\mathbf
e}_\omega . \]
 In the
following, $o_n(\varrho)$ with $n \in N$, denotes  a smooth
function of the order of  $\varrho$ such that $\lim_{\varrho
\rightarrow \,0}\ o_n(\varrho) = 0$. Similarly,
 \[
\nabla^2  \Psi = {\rm div\ grad (\Psi)} =\frac{1}{\varrho}\ h +
\frac{\partial h}{\partial \theta}\ \frac{d\theta}{d\varrho} +
\frac{1}{\varrho}\ \frac{\partial^2 h}{\partial \omega^2} +
\frac{1}{\varrho}\  o_{_4}(\varrho)= \frac{1}{\varrho}\ h +
\frac{1}{\varrho}\ \frac{\partial^2 h}{\partial \omega^2}
+\frac{1}{\varrho}\ o_{_5}(\varrho)  ,\] $\displaystyle {\rm
grad}(\nabla^2  \Psi )= \left(- \frac{1}{\varrho^2}\ (h +
\frac{\partial^2 h}{\partial \omega^2})  + \frac{1}{\varrho}\
(\frac{\partial h}{\partial \theta}  + \frac{\partial^3
h}{\partial \omega^2 \partial \theta})\frac{d\theta}{d\varrho}
+\frac{1}{\varrho^2}\ o_{_6}(\varrho) \right )\ {\mathbf
e}_\varrho\
\\ + \frac{1}{\varrho^2}\ (\frac{\partial h}{\partial \omega} +
\frac{\partial^3 h}{\partial \omega^3})\ {\mathbf e}_\omega  =
\left(- \frac{1}{\varrho^2}\ (h + \frac{\partial^2 h}{\partial
\omega^2}) + \frac{1}{\varrho^2}\ o_{_7}(\varrho) \right ){\mathbf
e}_\varrho +   \frac{1}{\varrho^2}\ (\frac{\partial h}{\partial
\omega}  + \frac{\partial^3 h}{\partial \omega^3})\ {\mathbf
e}_\omega , $\\

\noindent and finally,
\[ \nabla^4  \Psi =  \frac{1}{\varrho^3}\ (h +
\frac{\partial^2 h}{\partial \omega^2}) - \frac{1}{\varrho^2}\
(\frac{\partial h}{\partial \theta} + \frac{\partial^3 h}{\partial
\omega^2 \partial \theta})\frac{d\theta}{d\varrho} +
\frac{1}{\varrho^3}\
 (\frac{\partial^2 h}{\partial \omega^2}  + \frac{\partial^4
h}{\partial \omega^4})+ \frac{1}{\varrho^3}\ o_{_8}(\varrho)
\]
$\displaystyle =  \frac{1}{\varrho^3}\ (h + 2 \frac{\partial^2
h}{\partial \omega^2}  + \frac{\partial^4 h}{\partial \omega^4}) +
\frac{1}{\varrho^3}\ o_{_9}(\varrho) , $\\

\noindent The principal part of the Laurent expansion in
$\varrho\, $ of $\, \nabla^4 \Psi$ leads to the partial derivative
equation
$$
h + 2\ \frac{\partial^2 h}{\partial \omega^2}  + \frac{\partial^4
h}{\partial \omega^4} = 0 ,
$$
and  the general solution for the principal part of $\Psi$ is
$$
 \Psi (\varrho,\omega )=\varrho\ (a\ \mathrm{sin}\ \omega +b\
\mathrm{cos}\ \omega +c\ \omega \ \mathrm{sin}\ \omega +d\ \omega
\ \mathrm{cos}\ \omega), \eqno (21)
$$
where $a, b, c, d$ are functions of $\theta$. When $\mu_A =
\mu_B$, the boundary conditions at the solid wall and the
conditions $(i)$-$(iv)$ presented  in section 4  yield   the
values of coefficients $a, b, c, d$  for the two fluids $L_A$ and
$L_B$. These values are   given by relations $(13)_2$,  but here,
$\theta$ is not constant.
 The viscous stress
components are
 \[ \displaystyle\tau _{\varrho\omega
}=-\;\frac{\mu}{\varrho}(h+\frac{\partial^2 h}{\partial \omega^2}
)\ \equiv \ -\ 2\ \frac{ \mu }{\varrho}\ (c\ \mathrm{cos}\ \omega
-d\ \mathrm{sin}\ \omega ),\  \tau _{\varrho\varrho}=\tau
_{\omega\omega}=0.
\]
As in section 4, the pressure field is given by
\[
p=  p_{0}-2\ \frac{\mu }{\varrho}(c\ \mathrm{sin}\ \omega \ +d\
\mathrm{cos}\ \omega ) .
\]
In partial wetting, the stream function (21)  together with
relations $(13)_2$ show that   the partial derivatives of $h$ with
respect to $\omega$ and $\theta$ are bounded along $C_{t2}$. Along
$C_{t2}$, $p_A-p_B = \sigma_{AB}/R $ and consequently,
$$
\frac{\varrho\, d\theta }{dl} = \frac{\mu u}{\sigma_{AB}} \;2\pi\;
\mathrm{sin} \theta \; D(\theta). \eqno(22)
$$
Furthermore, $\|{\mathbf \Delta}\| = \{ 1/(2\varrho) )\}\; |\ h +
\partial^2 h/\partial \omega^2  |_{\omega=\theta}$. Thus,
\[
  \|{\mathbf \Delta}\| = \frac{|u|}{\varrho}\,|c_A \cos\,\theta -
  d_A\sin\,\theta|\equiv\frac{|u|}{\varrho}\,|c_B \cos\,\theta -
  d_B\sin\,\theta|\equiv  \frac{|u|}{\varrho}\, \sin^3\theta\,
  D(\theta) .
\]
Let us note that, in the frame $(O,{\mathbf n}^\prime_1, {\mathbf
n}_1)$, relations $ x =  \varrho\ {\rm cos}\, \theta $, $ y =
\varrho\ {\rm sin}\, \theta $, $ dy = dl\ {\rm sin}\, \theta$,
$\theta_i = {\mathrm {Arccos}}
\{(\sigma_{BS}-\sigma_{AS})/\sigma_{AB}\}$ together with Eq. (22)
and  relation (15),  allow us to obtain the parametric
representation of  $C_{t2}$ near the solid wall.

\noindent  Eq. (22) allows us  to verify that $\lim_{r \rightarrow
\,0\,\,} r\, d\theta/dl = 0$  and thus, the choice of the stream
function $\Psi$ in the form (21) together with relations $(13)_2$
is justified in the vicinity of the point $O$. Let us note that
when $u > 0$ (resp. $u < 0$), $\theta$ is an increasing (resp.
decreasing) function of the distance $y$ of a point of $C_{t2}$ to
the solid wall.
 We do notice also that, whereas the curvature of $C_{t2}$
tends to infinity when the point $O$ is approached, the stream
function $\Psi$ has the same form   as the stream function $\psi$
proposed in subsection 4.2 for a plane interface (a good example
of such a curve is given, near $x = 0$, by $y=\left| x\right|
^{\frac{3}{2} }$ (Voinov, \cite{Voinov})).
\subsection{ Apparent dynamic contact angle and  line friction}
Let $A$ be the point of $C_{t2}$ associated with  the cut-off
coefficient value $\gamma_0$, defined in subsection 4.2: the point
$A$ is at the border between the Newtonian and the non-Newtonian
domains of the fluid flows. We call {\it apparent dynamic contact
angle}   $\theta_a$, the value of $\theta$ associated with the
point $A$  (see fig. 4). Along the fluid-fluid interface,
condition $(Q_{Ai}^{j}-Q_{Bi}^{j})\ n_{2}^{i}n_{2j} = 0$, together
with relation  (12) and $ p_B-p_A = (1/\varrho)\;  \mu u  \;2\pi\;
\mathrm{sin} \theta \; D(\theta)$, imply
$$
\sigma _{AS}-\sigma _{BS}+\sigma _{AB}\;\mathrm{\cos \;}\theta
_{a}+\nu \;u\;=0,\eqno(23)
$$
with
$$
\nu = \int_{0}^{A} \frac{\mu}{\varrho}\;2\pi\;
\mathrm{sin}^{2}\theta \; D(\theta) dl. \eqno(24)
$$
Relation (23) is a form of \textit{Young-Dupr\'e dynamic relation
for the apparent dynamic contact angle}. We call $\nu$,
\textit{the line friction}. It is easy to verify that the scalar
$\nu$ is positive and of the same physical dimension  as a dynamic
viscosity. This result corresponds to the
 assumption in the article of Stokes {\it et al},
\cite{sto}, in which they say that \textit{there is an additional
viscous force on a moving contact line}. Other expressions for the
line friction have also been proposed (an attempt is done by a
thermodynamic point of view in Fan, Gao and Huang, \cite{fan}).

\noindent In the case of equilibrium, relation (23) yields the
static Young-Dupr\'e relation (Levitch, \cite{levitch})
\[
\sigma_{AS}-\sigma_{BS}+\sigma_{AB}\;\mathrm{\cos \;}\theta
_{e}=0,
\]
in which $\theta_e$ is the balance Young angle and $\theta_e =
 \theta_i =  \theta_a$.
 For any value of the contact line celerity, relations
(23) and (24) yield, implicitly, the apparent dynamic contact
angle $\theta_a$. With the formula (23), a simple explanation of a
well-known experimental result  (Dussan, \cite{dussan2}), may also
be corroborated: with the advance of the contact line, $u$ is
positive and the apparent dynamic contact angle $\theta_a$ is
larger than the equilibrium angle $\theta_e$. This result is
reversed when $u$ is negative.
\subsection{ Numerical investigations of the apparent dynamic contact angle and the line friction}
Hoffman, \cite{hoffman}, Legait and Sourieau,  \cite{legait},
Ram\'{e} and Garoff, \cite{rame},  and many other authors
experimentally observe that, near the contact line, for slow
motions, {\it the apparent dynamic contact angle  seems
independent of the microscopic distance to the solid surface}.
Let us verify numerically this observation.

\noindent   Using the relations $y = \varrho\; \mathrm{sin}\;
\theta$ and $dy = dl\; \mathrm{sin}\; \theta$, relation (22)
implies
\[ \frac{dy}{y} = \frac{\sigma_{AB}}{\mu\; u\; 2\pi\; \mathrm{sin}\; \theta
\; D(\theta)}\ d\theta .
\]
Let us consider a point $N$ of $C_{t2}$ in the Newtonian domain of
the fluid flows. Then,
 $$\ln
(\frac{y_n}{y_a}) = \int_{\theta_a}^{\theta_n}
\frac{\sigma_{AB}}{\mu_0\;  u\; 2\pi\; \mathrm{sin}\; \theta \;
D(\theta)}\ d\theta, \eqno (25)
$$
where $y_a$ and $y_n$ denote the distances of  points $A$ and $N$
to the solid wall. In partial wetting, when $\pi/6 < \theta <
5\pi/6$, then $0 < 2\pi\; \mathrm{sin}\; \theta \; D(\theta)  <
15$ and if $u > 0$,
$$
0 < \theta_n - \theta_a < 15\ C_a\ \ln(\frac{y_n}{y_a}).
$$
If we consider the case when $y_n = 10^4\ y_a $, a crude
approximation yields $ \theta_n - \theta_a   < 140\ C_a$ and thus
$\theta_n - \theta_a $ tends to zero with $C_a$. For example, when
$C_a = 10^{-4}$, we obtain  $\theta_n - \theta_a < 0.014$ radian,
(i.e. $0.8$ degree), and the apparent dynamic contact angle seems
 independent  of the distance of the point $N$ to
the solid wall: in the lubrication approximation for two
dimensional flows,   Eq. (23) expresses the behaviour of the
apparent dynamic contact angle independently of any microscopic
distance to the contact line. This result is in accordance with
  Seppecher's calculations  \cite{seppecher}.
\\

\noindent Let us estimate an order of magnitude of the line
friction.
 Along  $C_{t2}$, Eq. (21) implies that, for each fluid, $|\, h
+
\partial^2 h/\partial \omega^2  |_{_{\omega=\theta}} = |u|\, 2\, \mathrm{sin}^{3}\theta
\,D(\theta)$. When $\xi = 0.1$, a numerical computation yields
\[
\frac{\pi}{6} < \theta <\frac{5\;\pi}{6}\ \ \Rightarrow \ \ 1.015\
|u|^\xi  < |\ h + \frac{\partial^2 h}{\partial \omega^2}\
|_{_{\omega=\theta}} ^\xi < 1.032\ |u|^\xi.
\]
Taking into account  the norm $\|\mathbf \Delta\|$ of the strain
rate tensor along $C_{t2}$, relation (15) yields  a value of
$\gamma$ such that
\[
\gamma \simeq \frac{1}{1.02}\Big (\frac{2\ \varrho}{\tau_0 u}\Big
)^ \xi,
 \]
and $d\gamma/\gamma = \xi\ d\varrho/\varrho$. Then, relation (24)
allows us to obtain the value of the line friction
\[
\nu =\frac{2\pi \mu _{0}}{\xi }\,
\int_{0}^{\gamma_0}\mathrm{sin}^{2}\theta\;D(\theta)\
\frac{1-e^{-\gamma }}{\gamma }\ d\gamma.
\]
Since $\mathrm{sin}^{2}\theta \,D(\theta) > 0$, we obtain
\[
\nu =\frac{2\pi \mu _{0}}{\xi }\  \mathrm{sin}^{2}\theta
_{r}\;D(\theta _{r})\int_{0}^{\gamma _{0}}\frac{1-e^{-\gamma
}}{\gamma }d\gamma ,\ \ {\rm with} \ \ \theta_r \in [\;\theta_i,
\theta_a\;]
\]
in which $\theta_r$  is a convenient angle.  Due to $
\displaystyle \int_{0}^{\gamma _0 } (1/\gamma)(1-e^{-\gamma})
d\gamma \simeq 1 , $
 we obtain
$$
\nu =\frac{2\pi \mu _{0}}{\xi }\  \mathrm{sin}^{2}\theta
_{r}\;D(\theta _{r}) .
$$
In partial wetting, a numerical computation implies
\[
\frac{\pi}{6} < \theta_r <\frac{5\;\pi}{6}\ \ \Rightarrow \ \
 0.68 < \mathrm{sin}^{2}\theta
_{r} \;D(\theta _{r}) < 1.17 ,
\]
and consequently $$ 42\ \mu_0 < \nu < 73\ \mu_0 . \eqno (26) $$
Eq. (10) and Eq. (23) imply $|\cos\, \theta_a - \cos\, \theta_i| <
\nu\, |u| /\sigma_{AB}$. Taking into account inequalities (26), we
obtain $|\cos\, \theta_a - \cos\, \theta_i| <  73\, C_a$. In the
partial wetting case, for  $\theta \in (\pi/6,5\, \pi/6)$, then
$|\, \sin\big (\,(\theta_a -  \theta_i)/2\,\big)\,| <  73\, C_a$
and the derivative of $\sin^{2}\theta \;D(\theta)$ belongs to
$(-1,1)$; we can deduce  $|\sin^{2}\,\theta _{r} \,D(\theta _{r})
- \sin^{2}\theta_{a} \,D(\theta _{a})| < 2\, {\rm Arcsin}(73\,
C_a) $. This crude approximation proves that for $C_a$
sufficiently small, $\sin^{2}\theta_{r} \;D(\theta _{r})$ is close
to $\sin^{2}\theta _{a} \;D(\theta _{a})$ (for example, when $C_a
= 10^{-4}$, we obtain $|\sin^{2}\,\theta _{r} \,D(\theta _{r}) -
\sin^{2}\theta_{a} \,D(\theta _{a})| < 0.015$). Replacing
$D(\theta _{a})$ by its explicit expression sets the approximative
relation of the line friction :
$$ \nu =  \frac{2\pi
\mu _{0}}{\xi }\  \Big ( \; \frac{ \mathrm{sin}^{2}\,\theta_a}{
\theta_a \ (\pi -\theta_a ) - (\pi -2\; \theta_a)\ \mathrm{sin} \
\theta_a\ \mathrm{cos}\ \theta_a - \mathrm{sin}^2\ \theta_a }\;
\Big) . \eqno (27)
$$

\section{Concluding remarks}

In this paper, the dynamical problem of the contact of two
non-Newtonian viscous fluids $L_A$ and $L_B$ with a solid was
analyzed. Except at the contact line, these fluids were assumed to
adhere to the solid and to each other. Then, the principle of
virtual work allows us to obtain the governing equations and
boundary conditions. It was shown that the equations of motions
and boundary conditions lead to streamlines near the contact line
similar to those of a Newtonian fluid endowed with a dynamic
viscosity which is, when the strain rate tensor tends to zero, the
limit of the dynamic viscosity of the non-Newtonian fluid. The
analyze of the stress tensor and the dissipative function near the
contact line leads to several remarks and conclusions:\bigskip

 For dissipative movements, a viscous stress tensor
$Q_i^j$ was added to the pressure term and in the expression of
virtual work, the dissipative terms were distributed within the
volume as $Q_{i,j}^j$ and on the surfaces as $- Q_i^j n_{\alpha
j}$. In magnitude, the surface tension is comparable with the bulk
stress despite the fact that the thickness of the interfacial
layer, where the surface tension acts, is negligible compared to
the characteristic length scale in the bulk: although the
interfacial layer is very thin, the intermolecular forces which
act on it and give rise to the surface tension are very strong so
that the result is finite. Huge variations of the fluid velocity
appear at the three-phase contact line. The only physical factor,
which achieves to \textit{magnify} its role, does not come from
intermolecular forces but from the \textit{discontinuity} of the
velocity at the contact line and consequently from the viscosity
along the contact line when the thickness of the contact line
region tends to zero. In the immediate vicinity of the contact
line, the viscous stresses yield a friction force which acts on
the fluid-fluid interface and is balanced by strong capillary
tensions associated with the curvature of the fluid-fluid
interface. Consequently, expression (23) introduces a new term
associated with the contact line $\Gamma_t$.\bigskip

Creeping flow of a Newtonian fluid is unrealistic at the corner of
a moving contact line. Nevertheless, the system that is based on
Eq. (4), boundary conditions (5), (6), adherence assumption on
solid surfaces together with the dynamic Young-Dupr\'{e} relation
(23) for the apparent contact angle which \textit{accommodates the
behaviour of the\ non-Newtonian fluid domain} in the immediate
vicinity of the contact line poses a problem of slow fluid motion.
This result, in agreement with experiments and molecular
investigations, is brought out by means of the analytic
representation of the streamlines on Newtonian behaviour;
\textit{however, other non-Newtonian fluid behaviour is used to
arrive at  bounded dissipative functions near the contact line.}
\bigskip

For slow movements we are able to give a model which provides
answers to the previous questions:

- The contact line is a non material line and acts in a similar
way as a \textit{shock line}.

- The velocity fields are multivalued on the line.

- The paradox of infinite viscous dissipation is removed.

- Adherence and boundary conditions on surfaces and interfaces are
preserved, but a dynamic Young-Dupr\'{e} relation derived by the
virtual work principle yields the apparent dynamic contact angle
as an implicit function of the contact line celerity. The apparent
dynamic contact angle is the only pertinent Young angle from a
continuum mechanics point of view.

- For partial wetting, and for a sufficiently small capillary
number, the concept of line friction is associated with the
apparent contact angle.\bigskip

The contact angle hysteresis phenomenon and the modelling of
experimentally well-known results that express the dependence of
the dynamic contact angle on the celerity of the line are
important phenomena. In part 2,  Eq. (23) allows us to obtain an
explanation of the contact-angle hysteresis in the advance and
retreat of slowly moving fluids on a solid surface.

\section*{Acknowledgments}
I am grateful to Professor Seppecher and Professor Teshukov for
their helpful discussions about theoretical developments. I am
indebted to Professor Hutter and the anonymous referees for much
valuable criticism during the review process.

\end{document}